\documentclass[12pt]{article}
\def\text#1{\mathop{\rm #1}}
\def\lta{\mathrel{\displaystyle\mathop{\kern 0pt <}_{\raise .3ex
\hbox{$\sim$}}}}
\def\gta{\mathrel{\displaystyle\mathop{\kern 0pt >}_{\raise .3ex
\hbox{$\sim$}}}}
\input prepictex
\input pictex
\input postpictex
\newdimen\tdim
\tdim=\unitlength
\def\stpltsmbl{\setplotsymbol ({\small .})}
\def\tarrow{\arrow <5\tdim> [.3,.6]}

\begin{document}
\begin{flushright}
\setlength{\baselineskip}{3ex}
\#HUTP-97/A057\\ 9/97\end{flushright}

\begin{center}\Large {\bf {Sum-rule for large-$N_c$ QCD and
application to heavy quarkonia\footnote{Research supported in
part by the National Science Foundation under Grant \#PHY-9218167.}}\\
\vspace{12pt} }{\normalsize {\rm Dean Lee\footnote{{Supported in part by the
Fannie and John Hertz Foundation.}} and Howard
Georgi\\ \vspace{10pt} Harvard University\\ Cambridge, MA 02138\\
\vspace{12pt} 
\parbox{360pt}
{We introduce a new sum-rule for large-$N_c$ QCD which relates the density of
heavy quarkonium states, the state-averaged square of the wavefunction at the
origin, and the heavy quark current-current correlator. Focusing on the region
of energy just above perturbative threshold, we calculate the correlator by
incorporating  arbitrarily high orders in the QCD coupling $\alpha_s$.  We use
the sum-rule to determine the bottomonium potential using experimentally
measured s-wave leptonic widths and compare the result with the potential
obtained by direct calculation from the measured $s$-wave spectrum. We discuss
the utility of the sum-rule method for accurate determination of the confining
potential.}\\ \vspace{12pt} }} 
\end{center}

\section{Introduction}

\setcounter{footnote}{0}

In the heavy quarkonium system, we can study quark confinement without the
added complexity of relativistic processes. As the mass of the quark becomes
large, the dynamics of the quarkonium system becomes non-relativistic and
can be described by the Schr\"{o}dinger equation with a static potential.
The short distance behavior of this potential can be approximated by single
gluon exchange and therefore has the familiar Coulombic form. In this paper,
we consider the remaining long distance part of the potential, which we
refer to as the confining potential. In the course of our analysis we answer
the following questions: (i) Can the confining potential in some region of $r
$ be determined from the spectrum by direct calculation, i.e., without
parameter fits? (ii) Can the confining potential also be determined from
other data, such as the s-wave leptonic widths $\Gamma _{e^{+}e^{-}}$? (iii)
How do the results of (i) and (ii) compare, and can we accommodate both data
sets with one potential?

To answer these questions we make use of the large-$N_c$ expansion, where
$N_c$ is the number colors. We derive a sum-rule that relates the imaginary
part of the heavy quark current-current correlator with the product of $\rho 
$, the density of s-wave states, and $\left| \psi ^{ave}(0)\right| ^2,$ the
state-averaged square of the wavefunction at the origin. The derivation
involves calculating the imaginary part of an integral of the heavy quark
current-current correlator in momentum space. We do this calculation in two
different ways, once using perturbative methods and then again using
spectral sums. The sum-rule follows from equating the two calculations.

Sum-rules for the current-current correlator for light quark systems have a
long history~\cite{history}. Our approach is new in that we restrict our
attention to heavy quark currents and focus on the energy region just above
the perturbative $q\bar{q}$ threshold. As it turns out this energy region
has some interesting characteristics. One feature is the appearance of
threshold singularities in the perturbative calculation. To account for
these singularities, our calculation will include contributions from
arbitrarily large orders in $\alpha _s$, unlike more conventional treatments.
Another feature is that the effects of confinement can be described using
non-relativistic bound state methods. As we attempt to illustrate,
application of the sum-rule provides a useful complement to non-relativistic
methods. In particular, we discuss its use in the determination of
the confining potential for the bottomonium system.

The organization of the paper is as follows. In the sections 2 and 3, we
derive the sum-rule. In section 2, we discuss the perturbative calculation
of the current-current correlator, and we note that it is straightforward
to include information from arbitrarily high orders in $\alpha_s$ by using
the energy spectrum of the Coulomb bound states. In section 3, we calculate
the current-current correlator as a spectral sum and relate it to the
properties of the physical $q\bar q$ states in the large-$N_c$ limit. In
section
4, we test the sum-rule for the bottomonium system. In the remaining
sections we discuss various ways to determine the confining potential, discuss
our various approximations and the physics we have left out, and
answer the questions (i), (ii), and (iii).

\section{Perturbative calculation}

We consider quantum chromodynamics with $N_c$ colors and one flavor of heavy
quark, which we call $q$. Let $m$ be the mass of the heavy quark and $J^\mu
(x)$ be the corresponding electric current. We can write the renormalized
current-current correlator in momentum-space as 
\begin{equation}
\int\,d^4x\,e^{ikx}\,
\left\langle 0\right| T\left[ J^\mu (x)J^\nu (0)\right] \left|
0\right\rangle =\lim_{\varepsilon \rightarrow 0^{+}}\left( g^{\mu \nu
}-\frac{k^\mu k^\nu }{k^2+i\varepsilon }\right) \Pi (\sqrt{k^2},\varepsilon ),
\label{ndrt}
\end{equation}
where $\varepsilon $ is the infinitesimal positive parameter used to
regulate the singularities of the bare propagators. Because we are in the
time-like momentum region $k^2>0,$ $k^0>0$, we can move to the rest frame
where $\vec{k}=0$ and rewrite $\Pi (\sqrt{k^2},\varepsilon )$ as $\Pi
(k^0,\varepsilon )$.

Let $z$ and $\Delta z$ be fixed positive numbers such that $\Delta z<<z<<1$.
Let us define the integral 
\begin{equation}
P(m,z,\Delta z)=\lim_{\varepsilon \rightarrow 0^{+}}\int\limits_{\zeta \in
C(z,\Delta z)}\Pi\Bigl((2+\zeta )m,\varepsilon \Bigr)\,d\zeta ,  \label{ee}
\end{equation}
where $C$ is the contour in the complex $\zeta$-plane shown in Figure
1.\footnote{To avoid the possibility of endpoint singularities we have pulled
the
endpoints of the contour slightly off the real axis and into the upper half
plane, i.e, we have added a small positive imaginary part to $z$. This
technical point is of little consequence, however, since the limit
$\mathop{\rm Im}z\rightarrow 0^{+}$ presents no problems.} The purpose of
this contour integral will be explained in a moment when we discuss the
$N_c\rightarrow \infty $ limit. Our choice of energy values places us within
the non-relativistic $\bar{q}q$ resonance spectrum, just above the
perturbative threshold, $k^0=2m$.

If the dimensionless quantities $z$ and $\Delta z$ are held fixed, we expect 
$P$ to have a well-defined asymptotic expansion in $\alpha _s(m^2)$ as
$m\rightarrow \infty $, where $\alpha _s(m^2)$ is the QCD coupling\footnote{As
we will discuss section 6, the precise scale at which $\alpha_s$ is evaluated
does not matter to the order we are working. The difference between
$\alpha_s(m^2)$ and $\alpha_s(zm^2)$ is not important.} at
scale $m^2$. In the following we abbreviate $\alpha _s(m^2)$ as $\alpha _s$.
We are interested in the imaginary (absorptive) part of $P$, which we write
as 
\begin{equation}
\mathop{\rm Im}P(m,z,\Delta z)=N_c\frac{m^2}{2\pi }\Delta z\cdot G(z,\alpha
_s).  \label{rr}
\end{equation}
The perturbative expansion of $G$ will be discussed below. 
Let us now divide out the explicit $N_c$ in (\ref{rr})
and consider 
\begin{equation}
N_c^{-1}\mathop{\rm Im}P(m,z,\Delta z)  \label{erwrt}
\end{equation}
in the simultaneous limit $N_c\rightarrow \infty ,$ $m\rightarrow \infty $.
We rescale the coupling constant in the standard way so that $\alpha _sC_F$
has a well-defined limit as $N_c\rightarrow \infty $. We then have 
\begin{equation}
\lim_{N_c\rightarrow \infty }\left[ N_c^{-1}\mathop{\rm Im}P(m,z,\Delta
z)\right] =\frac{m^2}{2\pi }\Delta z\cdot G(z,\alpha _s).  \label{tedx}
\end{equation}

As $N_c\rightarrow \infty $, the $\bar{q}q$ resonances become stable and the
corresponding pole singularities approach the real $\zeta$-axis from below. As
$m\rightarrow \infty $, we expect the density of such pole singularities
approaching the real $\zeta$-axis to increase\footnote{This is discussed in
the
next section.} and anticipate that the integrand in
(\ref{ee}) will exist in the double limit $N_c\rightarrow \infty $,
$m\rightarrow \infty $ only in the sense of a distribution. The contour
integral defining $P$, however, does not have such problems. The pole
singularities approach the real axis from one side and therefore do not
pinch the contour. This is the motivation for our definition of $P$.

For $\alpha_s\ll z^{1/2},$ we can compute $G$ in (\ref{tedx}) 
by expanding in powers of $\alpha _s$,
i.e., the usual perturbative loop expansion. This expansion, however, is
inadequate for $z^{1/2}\lta \alpha _s.$ As the order of $\alpha_s $ increases,
the order of the $z^{-1/2}$ singularity also increases~\cite{appelquist}. It
is therefore more useful to expand $G$ as
\begin{equation}
G(z,\alpha _s)=\sum\limits_{n=1}^\infty (\alpha _sC_F)^n\cdot G_n(\alpha
_sC_Fz^{-1/2}),  \label{tr}
\end{equation}
where $C_F=\frac{N_c^2-1}{2N_c}.$ The leading term, $G_1$, is given by the
sum of all simple ladder diagrams, which in turn can be obtained from the
bound-state spectrum for the Coulomb potential. $G_1$ has been calculated
explicitly in \cite{fadin} (see also \cite{mbv}), with the result
\begin{equation}
G_1(x)=\frac 1x+\frac \pi 2+\frac x2\sum\limits_{n=1}^\infty \frac 1{
n^2+\frac{x^2}{4}}.  \label{vs}
\end{equation}
We will use this form in section 4.

\section{Spectral sum calculation}

Let us now repeat the calculation in (\ref{tedx}), this time using the
spectral representation of the current-current correlator. We consider the
contributions of all single and multiparticle states to the correlator: 
\begin{equation}
\left\langle 0\right| T\left[ J^\mu (x)J^\nu (0)\right] \left|
0\right\rangle
\begin{array}[t]{l}\displaystyle
=\theta (x^0)\sum_n\left\langle 0\right| J^\mu (x)\left| n\right\rangle
\left\langle n\right| J^\nu (0)\left| 0\right\rangle \\ \displaystyle
+\,\theta (-x^0)\sum_n\left\langle 0\right| J^v(0)\left| n\right\rangle
\left\langle n\right| J^\mu (x)\left| 0\right\rangle \,.
\end{array}
\label{herrt}
\end{equation}
In the $N_c\rightarrow \infty $ limit, the coupling of the current to
multiparticle meson states is suppressed by explicit factors of $N_c^{-\frac
12}$. The contribution of these states to the spectral sum in (\ref{herrt})
is therefore suppressed by $N_c^{-1}$, and we ignore such states in the
$N_c\rightarrow \infty $ limit.

Having reduced the correlator to a spectral sum over $\bar{q}q$ resonance
states, we use this spectral sum to compute the imaginary part of $N_c^{-1}P$.
In the $N_c\rightarrow \infty $ limit it is not necessary to include all
resonance states. Resonances which do not approach the contour $C$ give an
imaginary contribution that is suppressed by $N_c^{-1}$.\footnote{The real
contribution, however, is not suppressed.} Let us use the index $r$
to label resonances which approach the contour as $N_c\rightarrow \infty $.
For each resonance $r,$ let $M_r$ be the mass, ${\it \Gamma }_r$ be the
width, and $f_rg_{\mu \nu }$ be the coupling with the current. We then have 
\begin{equation}
N_c^{-1}\mathop{\rm Im}P(m,z,\Delta z)= 
\sum\limits_r\mathop{\rm Im}\left[ \int\limits_{\zeta \in C(z,\Delta
z)}\frac{-N_c^{-1}\left| f_r\right| ^2d\zeta }{\Bigl((2+\zeta )m\Bigr)^2-
\Bigl(M_r-\frac
i2{\it \Gamma }_r\Bigr)^2}\right] +\mathop{\rm }O(N_c^{-1})\mathop{\rm .}
\label{rex}
\end{equation}
>From here on, we suppress writing the $O(N_c^{-1})$ error term. We can
compute the integral in terms of the residues of the poles which approach $C$,

\begin{equation}
N_c^{-1}\mathop{\rm Im}P(m,z,\Delta z)=\sum_r\frac{\pi N_c^{-1}\left|
f_r\right| ^2}{2(2+z)m^2}\mathop{\rm .}  \label{wefd}
\end{equation}
Since these resonances are non-relativistic bound states, we can use the
van-Royen--Weisskopf result, 
\begin{equation}
N_c^{-1}\left| f_r\right| ^2=4\left| \psi _r(0)\right| ^2M_r,  \label{hsst}
\end{equation}
where $\psi _r(0)$ is the Schr\"{o}dinger wavefunction of the resonance at
the origin. Only s-wave states couple to the photon in this limit.
Substituting (\ref{hsst}) into (\ref{wefd}), we get 
\begin{equation}
N_c^{-1}\mathop{\rm Im}P(m,z,\Delta z)=\frac{2\pi }m\sum_r\left| \psi
_r(0)\right| ^2\mathop{\rm .}  \label{esx}
\end{equation}
Comparing (\ref{tedx}) and (\ref{esx}), we find 
\begin{equation}
\frac{2\pi }m\sum_r\left| \psi _r(0)\right| ^2=\frac{m^2}{2\pi }\Delta
z\cdot G(z,\alpha _s)\mathop{\rm .}  \label{jdre}
\end{equation}
The spectral sum on the left side of (\ref{jdre}) is a discrete sum. It
therefore has intrinsic fluctuations of size $[\Delta n(zm)]^{-1}$, where
$\Delta n(zm)$ is the number of s-wave states between $(2+z)m$ and
$(2+z+\Delta z)m$. These fluctuations are not reproduced in the perturbative
calculation at any finite order in $\alpha _s$. This is what we expect from an
operator product analysis in which the fluctuations are associated with
matrix elements of higher dimension operators, proportional to powers of
$\Lambda _{{\rm QCD}}$~\cite{ope}. We conclude that $[\Delta n(zm)]^{-1}$ is
smaller than any power of $\alpha _s$ as $m\rightarrow \infty $.

Let $\rho (zm)$ be the density of s-wave states at energy $zm$. We can
rewrite
\begin{equation}
\frac 1m\sum_r\left| \psi _r(0)\right| ^2\rightarrow\rho (zm)\left| \psi
_{zm}^{\mathop{\rm ave}}(0)\right| ^2\Delta z
\end{equation}
where $\left| \psi_{zm}^{\mathop{\rm ave}}(0)\right|^2$ is the average of
$\left| \psi_r(0)\right|^2$ for resonances with energies near $zm$. We then
have 
\begin{equation}
2\pi \rho (zm)\left| \psi _{zm}^{\mathop{\rm ave}}(0)\right|
^2=\frac{m^2}{2\pi }G(z,\alpha _s),  \label{vbn}
\end{equation}
or 
\begin{equation}
\rho (zm)=\frac{m^2G(z,\alpha _s)}{4\pi ^2\left| \psi _{zm}^{\mathop{\rm
ave}}(0)\right| ^2}.  \label{uy}
\end{equation}

At this point, 
the careful reader may wonder whether we have done something inconsistent,
because we have argued in this section that the physical $q\bar q$ bound
states affect the imaginary part of the current-current correlator only if
they approach the contour $C$, while in section 2, we saw that the
perturbative calculation of the same imaginary part receives contributions
which appear to be associated with the Coulomb bound states below
perturbative threshold, not on the contour. The difference stems from the
presence of the continuum states in the perturbative calculation. The
$\sqrt z$ singularity from the continuum states can be distorted by the
far-away poles. Effectively, the contribution to the imaginary part of the
correlator in
(\ref{tedx}-\ref{vs}) is a product of the real part of the Coulomb
bound state poles
with the imaginary part from the continuum states.

\section{Bottomonium test\label{test}}

In this section we evaluate the right and left sides of the sum-rule (\ref
{uy}) for bottomonium states. We begin with the right side. 
For our purposes we use the approximation 
\begin{equation}
G(z,\alpha _s)\approx \alpha _sC_F\cdot G_1(\alpha _sC_Fz^{-1/2}).
\label{jy}
\end{equation}
where $G_1$ is defined in (\ref{vs}), with values $m_b=4600$ MeV and
$\alpha_s(m_b^2)C_F=0.28.$ The
state-averaged square of the wavefunction at the origin, $\left| \psi
_{zm}^{\mathop{\rm
ave}}(0)\right| ^2$, can be determined from the leptonic widths of the
s-wave states. The relation is 
\begin{equation}
\left| \psi _{zm}^{\mathop{\rm ave}}(0)\right| ^2=\frac{\Gamma
_{e+e-}M^2}{16\pi Q^2\alpha _{em}^2},  \label{exs}
\end{equation}
where $\Gamma _{e+e-}$ is the decay width into $e^{+}e^{-}$, $M$ is the mass
of the meson (i.e., $2m+zm$), and $\alpha _{em}$ is the electromagnetic
coupling.

Let us define $E=zm$. The left side of ($\ref{uy}$) can be obtained in the
following manner. We first define a discrete function $n(E)$ such that
$n(E_{1s})=1,$ $n(E_{2s})=2$, $\cdots $, where $E_{ks}$ is the energy of the
$k^{th}$ s-wave meson. We smoothly interpolate between the values $E_{ks}$,
thereby extending $n(E)$ to a differentiable function. We then differentiate 
$n(E)$ to get the density of states, which we call $\rho _1(E).$ We will be
working with several versions of the densities of states and it is necessary
to give them different names. Let us define $\rho _2(E)$ as the right side
of ($\ref{uy}$), 
\begin{equation}
\rho _2(E)=\frac{m^2G(z,\alpha _s)}{4\pi ^2\left| \psi _E^{\mathop{\rm
ave}}(0)\right| ^2}\mathop{\rm .}  \label{xve}
\end{equation}
In Figure 2 we have plotted $\rho _1(E)$ and $\rho _2(E)$ as functions of
the meson mass $2m+E$. The results indicate that the sum-rule is satisfied
rather well.

\section{Determination of the potential}

In this section we propose several new methods for calculating the
bottomonium confining potential. The sum-rule ($\ref{uy}$) gave us two
calculations for the density of s-wave states, $\rho _1$ and $\rho _2$. In
this section we use WKB methods to convert $\rho _1$ and $\rho _2$ into
their corresponding potentials.

Let us write the potential for the heavy quarkonium system as 
\begin{equation}
V(r)=-\frac{\alpha _sC_F}r+U(r).  \label{dfb}
\end{equation}
We now take a moment to discuss the expected behavior of $U(r).$ In section
4 we found that the number of s-wave states with energy less than $zm$
grows                                                                       
more rapidly than any inverse power of $\alpha_s$ as $m\rightarrow \infty$.
The number of nodes of the s-wave wavefunctions with energy $zm$ must
then
scale faster than any inverse power of $\alpha_s$. This implies that
$U(r)$
is slowly varying on distance scales of size $m^{-1}z^{-\frac 12}$, the
approximate distance between nodes. We
also expect that the magnitude of $U(r)$ is small near $r=0.$ This is
because the leading term in the perturbative calculation, $\alpha_sC_F\cdot
G_1(\alpha _sC_Fz^{-1/2})$, is produced by the continuous spectrum of the
Coulomb potential. The bound states of $V(r)$ must therefore approximate
the
continuous states of the Coulomb potential and the contribution from $U(r)$
near $r=0$ must be small.

For the purposes of calculating energy levels and wavefunctions, we can
ignore the change of $U(r)$ on length scales $\Delta r\lta m^{-1}$ and
approximate $V(r)$ by $V_I(r)$:

\begin{equation}
V_I(r)=-\frac{\alpha _sC_F}r+U(r+m^{-1}).  \label{reg}
\end{equation}
This approximation introduces an error of size $\frac 1{E\cdot
M}\frac{dU}{dr%
}.$

Since $U(m^{-1})\ll E\ll m$, we can subtract a constant term, $U(m^{-1}),$
from $V$ by a small redefinition of the heavy quark mass 
\begin{equation}
m\rightarrow m+\frac 12U(m^{-1})  \label{er}
\end{equation}
\begin{equation}
V_I\rightarrow V_{II}=V_I-U(m^{-1})\mathop{\rm .}  \label{ex}
\end{equation}
This shift introduces an error on the wavefunctions and masses of the meson
states of size $\frac 1mU(m^{-1})$. Combining (\ref{ex}) with (\ref{reg}),
we have 
\begin{equation}
V_{II}(r)=-\frac{\alpha _sC_F}r+W(r),  \label{bf}
\end{equation}
where 
\begin{equation}
W(r)=U(r+m^{-1})-U(m^{-1})\mathop{\rm .}  \label{zxx}
\end{equation}

We now apply the WKB approximation. In the WKB approximation the density of
states is given by

\begin{equation}
\rho (E)=\frac{\sqrt{m}}\pi
\int_0^{V_{II}^{-1}(E)}\frac{dy}{2\sqrt{E-V_{II}(y)}},  \label{yf}
\end{equation}
where the size of the error is $n(E)^{-1}$. The Coulomb term effects the
integrand of (\ref{yf}) only in the small region $0\leq y\lta E^{-1}.$ As a
result, the Coulomb term makes only a small contribution to $\rho (E)$ of
size $\alpha _s$. We ignore this contribution and approximate $\rho (E)$ as 
\begin{equation}
\rho (E)\approx \frac{\sqrt{m}}{2\pi
}\int_0^{W^{-1}(E)}\frac{dr}{\sqrt{E-W(r)}}\mathop{\rm .}  \label{vn}
\end{equation}
Let us define a function $F(x)$ implicitly as 
\begin{equation}
\frac 1{F(W(r))}=\frac{dW(r)}{dr}\mathop{\rm ,}  \label{df}
\end{equation}
where we have assumed that $W(r)$ is an increasing function of $r$. We then
have 
\begin{equation}
\rho (E)\approx \frac{\sqrt{m}}{2\pi
}\int_0^E\frac{F(Y)dY}{\sqrt{E-Y}}\mathop{\rm ,}  \label{nf}
\end{equation}
which we recognize as an Abel transform of $F$. We can invert the transform, 
\begin{equation}
F(E)=\frac 2{\sqrt{m}}\frac d{dE}\int_0^E\frac{\rho (Y)dY}{\sqrt{E-Y}}.
\label{xb}
\end{equation}
Combining (\ref{df}) and (\ref{xb}) and integrating, we find 
\begin{equation}
r(W)=\frac 2{\sqrt{m}}\int_0^W\frac{\rho (Y)dY}{\sqrt{W-Y}}.  \label{dx}
\end{equation}
This gives us the inverse of $W(r)$. Using $\rho _1$ and $\rho _2$ from the
previous section, we can calculate the corresponding potentials $W_1$ and
$W_2$. Doing this calculation, we find that $W_1$ and $W_2$ agree quite well.
The results are shown in Figure 3. Purely for comparison, we plot
$\alpha_sC_F/r$, the negative of the Coulomb potential, to emphasize that it
is not important compared to the confining potential in most of the region of
interest.

\section{Discussion}

The large-$N_c$ sum-rule provides a new perspective on the heavy quarkonium
confining potential by drawing an explicit relation between the density of
states and the wavefunction at the origin. As we have seen, the confining
potential can now be determined in two different ways. In this section we
discuss our approximations and
consider the question of how to use this new information to determine the
confining potential more accurately.

We have made two approximations in the derivation of our sum-rule, large $N_c$
and small $\alpha_s$. We do not believe that the large $N_c$ approximation
introduces a very serious error
in this analysis. It is used to argue that the
bound-state poles are close to the real axis. Because the states we actually
use
in our analysis are rather narrow, this is probably reasonable. By working to
first order in $\alpha_sC_F$ in (\ref{tr}), we are ignoring a number of
potentially important effects, for example the running of the coupling. Thus,
for
example, we cannot say whether we should evaluate the coupling at $m^2$, as we
have done, or at a smaller mass scale like $E^2=(zm)^2$. We are thus ignoring
terms proportional to powers of $\alpha_s\ln(1/z)$. We must do this for our
perturbation series to be consistent. Because we work to all orders in
$\alpha_s/\sqrt{z}$, we can go down to $z$ for which this is of order one, but
for $\alpha_s\ln(1/z)$ to be large requires still smaller $z\approx e^{-
B/\alpha_s}\approx \Lambda_{QCD}/m$, which implies $E=zm\approx\Lambda_{QCD}$.
We expect perturbation theory to break down in this region.

Our calculation of $W_1$ and $W_2$ was done
by means of integral transform methods, rather than parameter fitting. We
believe this carries certain advantages, such as calculational simplicity
and the elimination of initial biases for the behavior of the potential. Our
results for $W_1$ and $W_2$ appear to indicate a logarithmic or small
fractional power law behavior for the confining potential in the relevant
region for the observed bottomonium spectrum ($r\lta 10^{-2}$ MeV$^{-1}$).
There are several ways to use our results to produce a more accurate
representation of the confining potential. One approach is to use $W_1$ or
$W_2$ or some average of the two as an initial result and then use numerical
fitting to find the best possible correction. We expect this correction to
be quite small, however, since as we now show the initial result appears to
be quite accurate already.

We incorporate data for both the s-wave energy levels and leptonic widths in
the most simple way by constructing an average density of states, 
\begin{equation}
\rho _{\mathop{\rm ave}}(E)=\frac{\rho _1(E)+\rho _2(E)}2.  \label{zb}
\end{equation}
We then use ($\ref{dx}$) to find the corresponding potential for $\rho
_{\mathop{\rm ave}}(E)$. As before we have taken $m_b=4600$ MeV\thinspace and
$\alpha _s(m_b^2)C_F=0.28$. We now put the Coulomb term back in and calculate
the resulting spectrum and leptonic widths. The results are shown in Table 1.

\section{Summary}

In our analysis of the heavy quarkonium confining potential we have answered
the three questions posed in the introduction, and the answer to each
question is positive. We derived a sum-rule that relates the imaginary part
of the heavy quark current-current correlator with the product of the
density of s-wave states and the state-averaged square of the wavefunction
at the origin. As we have seen, application of the sum-rule provides a
useful complement to non-relativistic potential methods.

\section{Acknowledgments}

We are grateful to S. Coleman, A. Hoang, R. Marshall, A. Radyushkin and M. Voloshin for
helpful discussions and suggestions.

\newpage\ 

\begin{center}
Table 1

\[
\begin{tabular}{lllll}
Level & Mass (th.) & Mass (ex.) & $\Gamma _{e^{+}e^{-}}$ (th.) & $\Gamma
_{e^{+}e^{-}}$ (ex.) \\ 
1s & 9490 MeV & 9460 MeV & 1.3 keV & 1.3 keV \\ 
2s & 10040 & 10020 & .63 & .52 \\ 
3s & 10370 & 10360 & .44 & .47 \\ 
4s & 10600 & 10580 & .34 & .59 \\ 
5s & 10820 & 10860 & .34 & .31 \\ 
6s & 11000 & 11020 & .17 & .13 \\ 
2p & 9900 & 9890 &  &  \\ 
3p & 10260 & 10260 &  & 
\end{tabular}
\]
\end{center}

\section*{Figure Captions}

\noindent 
Figure 1. Plot of the integration contour $C$.

\noindent
Figure 2. Plot of the densities $\rho_1$ and $\rho_2$.

\noindent
Figure 3. Plot of the confining potentials $W_1$ and $W_2$.
The dotted line is $\alpha_sC_F/r$.

\newpage

$$\beginpicture
\setcoordinatesystem units <1.5\tdim,1.5\tdim>
\setplotarea x from -25 to 150, y from -25 to 120
\tarrow from 0 -25 to 0 75
\tarrow from -25 0 to 150 0
\putrule from 125 -4 to 125 4
\put {1} at 125 14
\put {${\rm Re}\,\zeta$} at 137 -12
\put {${\rm Im}\,\zeta$} [l] at 6 80
\circulararc 360 degrees from 10 0 center at 25 0
\circulararc 360 degrees from 50 75 center at 95 75
\plot 19.5 2 19.5 1.5 20 1 30 1 30.5 1.5 30.5 2 /
\setdashes
\plot 12 7 52 90 /
\plot 32 -13 120 37 /
\setsolid
\stpltsmbl
\plot 50 75 140 75 /
\plot 80 77 110 77 /
\tarrow from 92 77 to 96 77
\put {$C$} at 95 84
\putrule from 80 72 to 80 78
\putrule from 110 72 to 110 78
\put {$z$} at 80 65
\put {$z+\Delta z$} at 110 65
\circulararc 90 degrees from 110 77 center at 110 80
\circulararc -90 degrees from 80 77 center at 80 80
\plotheading{Figure 1}
\linethickness=0pt
\putrule from 0 -25 to 0 75
\putrule from -25 0 to 150 0
\endpicture$$

\centerline{Figure 1. Plot of the integration contour $C$.}

\newpage

$$\beginpicture
\setcoordinatesystem units <.15\tdim,170\tdim>
\setplotarea x from 9250 to 11250, y from 0 to 1.1
\inboundscheckon
\axis bottom visible ticks short in from 9250 to 11250 by 50 /
\axis bottom visible ticks in numbered from 9500 to 11000 by 250 /
\put {0.002} [r] at 9200 .2
\put {0.004} [r] at 9200 .4
\put {0.006} [r] at 9200 .6
\put {0.008} [r] at 9200 .8
\put {0.01} [r] at 9200 1
\put {0} [r] at 9200 0
\put {$\rho_1$} at 11100 .75
\put {$\rho_2$} at 10850 .9
\axis left visible ticks short in from 0 to 1.1 by .1 /
\axis left visible ticks in from .2 to 1 by .2 /
\plotheading{Figure 2}
\put {Mass (MeV)} at 10250 -.15
\put {$\rho$ (MeV$^{-1}$)} [r] at 9150 .5
\visibleaxes
\normalgraphs
\plot
9300 0.111152 9320 0.114163 9340 0.117174 9360 0.120185 
    9380 0.123196 9400 0.126207 9420 0.129218 9440 0.132229 
    9460 0.13524 9480 0.138251 9500 0.141262 9520 0.144273 
    9540 0.147284 9560 0.150295 9580 0.153306 9600 0.156317 
    9620 0.159328 9640 0.162339 9660 0.16535 9680 0.168361 
    9700 0.171372 9720 0.174383 9740 0.177394 9760 0.180405 
    9780 0.183416 9800 0.186427 9820 0.189438 9840 0.192449 
    9860 0.19546 9880 0.198471 9900 0.201482 9920 0.204493 
    9940 0.207504 9960 0.210515 9980 0.213526 10000 0.216537 
    10020 0.219548 10040 0.230435 10060 0.24226 10080 
    0.253597 10100 0.264446 10120 0.274807 10140 0.28468 
    10160 0.294065 10180 0.302963 10200 0.311372 10220 
    0.319293 10240 0.326727 10260 0.333672 10280 0.34013 
    10300 0.3461 10320 0.351581 10340 0.356575 10360 0.36972 
    10380 0.404486 10400 0.432667 10420 0.454263 10440 
    0.469276 10460 0.477704 10480 0.479547 10500 0.474807 
    10520 0.463481 10540 0.445572 10560 0.421078 10580 0.39 
    10600 0.369271 10620 0.352391 10640 0.339359 10660 
    0.330175 10680 0.32484 10700 0.323353 10720 0.325714 
    10740 0.331924 10760 0.341983 10780 0.355889 10800 
    0.373644 10820 0.395248 10840 0.4207 10860 0.45 10880 
    0.49375 10900 0.5375 10920 0.58125 10940 0.625 10960 
    0.66875 10980 0.7125 11000 0.75625 11020 0.8 11040 
    0.84375 11060 0.8875 11080 0.93125 11100 0.975 11120 
    1.01875 11140 1.0625 11160 1.10625 11180 1.15 11200 
    1.19375
/
\plot
9300 0.0971465 9320 0.0998681 9340 0.102548 9360 0.105177 
    9380 0.107818 9400 0.110474 9420 0.11317 9440 0.115923 
    9460 0.118745 9480 0.121651 9500 0.124651 9520 0.127758 
    9540 0.130982 9560 0.134336 9580 0.137832 9600 0.141484 
    9620 0.145304 9640 0.149309 9660 0.153516 9680 0.15794 
    9700 0.162605 9720 0.16753 9740 0.17274 9760 0.178268 
    9780 0.184134 9800 0.190386 9820 0.197055 9840 0.20419 
    9860 0.211852 9880 0.220077 9900 0.228965 9920 0.238577 
    9940 0.249014 9960 0.260419 9980 0.274527 10000 0.288001 
    10020 0.297898 10040 0.303713 10060 0.307377 10080 
    0.309722 10100 0.31121 10120 0.312708 10140 0.314508 
    10160 0.316319 10180 0.318142 10200 0.319979 10220 
    0.321829 10240 0.323694 10260 0.325572 10280 0.327299 
    10300 0.329233 10320 0.330755 10340 0.332867 10360 
    0.338184 10380 0.35127 10400 0.368028 10420 0.386293 
    10440 0.405546 10460 0.426919 10480 0.450522 10500 
    0.477078 10520 0.510035 10540 0.547002 10560 0.579744 
    10580 0.597991 10600 0.607055 10620 0.605672 10640 
    0.595813 10660 0.582677 10680 0.572996 10700 0.563609 
    10720 0.554501 10740 0.54566 10760 0.537075 10780 
    0.528733 10800 0.520626 10820 0.508691 10840 0.496971 
    10860 0.497613 10880 0.529487 10900 0.577599 10920 
    0.633252 10940 0.696703 10960 0.777167 10980 0.898402 
    11000 1.02438 11020 1.12713 11040 1.20062 11060 1.26068 
    11080 1.31196 11100 1.35989 11120 1.40966 11140 1.46645 
    11160 1.53545 11180 1.62183 11200 1.73079
/
\endpicture$$

\centerline{Figure 2. Plot of the densities $\rho_1$ and $\rho_2$.}

\newpage

$$\beginpicture
\setcoordinatesystem units <180\tdim,.09\tdim>
\setplotarea x from 0 to 1.75, y from 0 to 2200
\axis bottom visible ticks short in from 0 to 001.75 by 000.05 /
\axis bottom visible ticks in from 000.25 to 001.50 by 000.25 /
\put {$0$} at 0 -100
\put {$0.0025$} at 0.25 -100
\put {$0.005$} at 0.5 -100
\put {$0.0075$} at 0.75 -100
\put {$0.01$} at 1.0 -100
\put {$0.0125$} at 1.25 -100
\put {$0.015$} at 1.5 -100
\put {$W_1$} at 000.65 1800
\put {$W_2$} at 000.95 1500
\axis left visible ticks short in from 100 to 2100 by 100 /
\axis left visible ticks in numbered from 0 to 2000 by 500 /
\plotheading{Figure 3}
\put {Radius (MeV$^{-1}$)} at 0.875 -250
\put {$W$ (MeV)} [r] at -0.07 1250
\visibleaxes
\normalgraphs
\plot
0.01 15.6736 0.02 29.3471 0.03 43.0207 0.04 60.6182 0.05 
    85.7213 0.06 112.197 0.07 141.205 0.08 171.421 0.09 
    202.363 0.1 234.249 0.11 266.319 0.12 298.619 0.13 
    331.003 0.14 363.36 0.15 395.648 0.16 427.752 0.17 
    459.736 0.18 491.488 0.19 523. 0.2 554.366 0.21 585.375 
    0.22 616.198 0.23 646.805 0.24 677.073 0.25 707.2 0.26 
    736.989 0.27 766.575 0.28 795.957 0.29 822.575 0.3 
    848.478 0.31 868.957 0.32 888.582 0.33 907.578 0.34 
    925.213 0.35 942.849 0.36 960.062 0.37 976.818 0.38 
    993.575 0.39 1010.16 0.4 1026.56 0.41 1042.97 0.42 
    1059.37 0.43 1075.77 0.44 1092.17 0.45 1108.67 0.46 
    1125.34 0.47 1142. 0.48 1156.55 0.49 1167.93 0.5 
    1179.31 0.51 1190.69 0.52 1202.05 0.53 1211.82 0.54 
    1221.59 0.55 1231.36 0.56 1241.13 0.57 1250.9 0.58 
    1262.25 0.59 1273.81 0.6 1285.37 0.61 1296.92 0.62 
    1312.3 0.63 1330.67 0.64 1349.03 0.65 1442.44 0.66 
    1532.2 0.67 1566. 0.68 1589.75 0.69 1608.94 0.7 1623.27 
    0.71 1637.6 0.72 1651.93 0.73 1660.98 0.74 1670. 0.75 
    1679.02 0.76 1688.05 0.77 1697.07 0.78 1705.02 0.79 
    1711.68 0.8 1718.34 0.81 1725.01 0.82 1731.67 0.83 
    1738.33 0.84 1744.99 0.85 1751.65 0.86 1757.32 0.87 
    1762.92 0.88 1768.53 0.89 1774.14 0.9 1779.75 0.91 
    1785.36 0.92 1790.97 0.93 1796.58 0.94 1802.16 0.95 
    1807.12 0.96 1812.07 0.97 1817.02 0.98 1821.98 0.99 
    1826.93 1. 1831.88 1.01 1836.84 1.02 1841.79 1.03 
    1846.74 1.04 1851.7 1.05 1856.22 1.06 1860.71 1.07 
    1865.2 1.08 1869.69 1.09 1874.18 1.1 1878.67 1.11 
    1883.16 1.12 1887.66 1.13 1892.15 1.14 1896.64 1.15 
    1901.13 1.16 1905.34 1.17 1909.48 1.18 1913.62 1.19 
    1917.77 1.2 1921.91 1.21 1926.05 1.22 1930.19 1.23 
    1934.34 1.24 1938.48 1.25 1942.62 1.26 1946.76 1.27 
    1950.9 1.28 1955.05 1.29 1959.19 1.3 1963.33 1.31 
    1967.47 1.32 1971.62 1.33 1975.76 1.34 1979.9 1.35 
    1984.04 1.36 1988.18 1.37 1992.33 1.38 1996.47 1.39 
    2000.61 1.4 2004.75 1.41 2008.9 1.42 2013.04 1.43 
    2017.18 1.44 2021.32 1.45 2025.46 1.46 2029.61 1.47 
    2033.75 1.48 2037.89 1.49 2042.03 1.5 2046.18 1.51 
    2050.32 1.52 2054.46 1.53 2058.6 1.54 2062.75 1.55 
    2066.89 1.56 2071.03 1.57 2075.17 1.58 2079.31 1.59 
    2083.46 1.6 2087.6 1.61 2091.74 1.62 2095.88 1.63 
    2100.03 1.64 2104.17 1.65 2108.31 1.66 2112.45 1.67 
    2116.59 1.68 2120.74 1.69 2124.88 1.7 2129.02
/
\plot
0.01 11.5514 0.02 27.7376 0.03 49.5418 0.04 76.0303 0.05 
    105.922 0.06 138.436 0.07 172.795 0.08 208.34 0.09 
    244.495 0.1 280.779 0.11 316.797 0.12 352.232 0.13 
    386.823 0.14 420.406 0.15 452.852 0.16 484.063 0.17 
    514.006 0.18 542.655 0.19 570.009 0.2 596.088 0.21 
    620.918 0.22 644.541 0.23 666.998 0.24 688.336 0.25 
    708.616 0.26 727.89 0.27 746.161 0.28 763.214 0.29 
    779.321 0.3 794.876 0.31 810.155 0.32 825.436 0.33 
    840.993 0.34 857.233 0.35 874.466 0.36 892.393 0.37 
    911.061 0.38 930.414 0.39 950.286 0.4 970.519 0.41 
    991.12 0.42 1012.05 0.43 1033.28 0.44 1054.8 0.45 
    1076.59 0.46 1098.61 0.47 1121.29 0.48 1143.57 0.49 
    1163.91 0.5 1182.13 0.51 1198.73 0.52 1213.72 0.53 
    1227.34 0.54 1239.85 0.55 1251.44 0.56 1262.85 0.57 
    1273.7 0.58 1283.99 0.59 1293.76 0.6 1303.03 0.61 
    1311.74 0.62 1320.06 0.63 1328.05 0.64 1335.76 0.65 
    1343.26 0.66 1350.61 0.67 1357.77 0.68 1364.89 0.69 
    1372.07 0.7 1379.39 0.71 1386.92 0.72 1394.76 0.73 
    1402.88 0.74 1411.05 0.75 1420.04 0.76 1429.9 0.77 
    1440.65 0.78 1452.34 0.79 1465. 0.8 1478.67 0.81 
    1493.38 0.82 1509.1 0.83 1525.92 0.84 1544.05 0.85 
    1563.19 0.86 1584.14 0.87 1608.83 0.88 1637.71 0.89 
    1665.05 0.9 1687.29 0.91 1700.9 0.92 1702.33 0.93 
    1707.96 0.94 1712.59 0.95 1716.44 0.96 1720.48 0.97 
    1725.66 0.98 1732.95 0.99 1743.3 1. 1753.16 1.01 
    1756.07 1.02 1758.63 1.03 1760.97 1.04 1763.21 1.05 
    1765.5 1.06 1767.95 1.07 1770.71 1.08 1773.9 1.09 
    1777.66 1.1 1782.1 1.11 1787.38 1.12 1793.61 1.13 
    1800.93 1.14 1804.75 1.15 1807.89 1.16 1810.97 1.17 
    1814. 1.18 1816.99 1.19 1819.94 1.2 1822.85 1.21 
    1825.74 1.22 1828.59 1.23 1831.42 1.24 1834.23 1.25 
    1837.02 1.26 1839.8 1.27 1842.57 1.28 1845.33 1.29 
    1848.09 1.3 1850.85 1.31 1853.74 1.32 1856.71 1.33 
    1859.69 1.34 1862.67 1.35 1865.66 1.36 1868.65 1.37 
    1871.64 1.38 1874.64 1.39 1877.64 1.4 1880.63 1.41 
    1883.64 1.42 1886.64 1.43 1889.64 1.44 1892.64 1.45 
    1895.64 1.46 1898.63 1.47 1901.63 1.48 1904.62 1.49 
    1907.61 1.5 1910.6 1.51 1913.58 1.52 1916.56 1.53 
    1919.54 1.54 1922.5 1.55 1925.47 1.56 1928.42 1.57 
    1931.37 1.58 1934.31 1.59 1937.24 1.6 1940.17 1.61 
    1943.08 1.62 1945.99 1.63 1948.89 1.64 1951.77 1.65 
    1954.65 1.66 1957.51 1.67 1960.36 1.68 1963.2 1.69 
    1966.03 1.7 1968.84
/
\setdots <2\tdim>
\plot
 .0127  2200 
 .013   2154 
 .014   2000 
 .015   1867 
 .016   1750 
 .017   1647 
 .018   1556 
 .019   1474 
 .02   1400 
 .021   1333 
 .022   1273 
 .023   1217 
 .024   1167 
 .025   1120 
 .026   1077 
 .027   1037 
 .028   1000 
 .029   966 
 .03   933 
 .031   903 
 .032   875 
 .033   848 
 .034   824 
 .035   800 
 .036   778 
 .037   757 
 .038   737 
 .039   718 
 .04   700 
 .041   683 
 .042   667 
 .043   651 
 .044   636 
 .045   622 
 .046   609 
 .047   596 
 .048   583 
 .049   571 
 .05   560 
 .051   549 
 .052   538 
 .053   528 
 .054   519 
 .055   509 
 .056   500 
 .057   491 
 .058   483 
 .059   475 
 .06   467 
 .061   459 
 .062   452 
 .063   444 
 .064   438 
 .065   431 
 .066   424 
 .067   418 
 .068   412 
 .069   406 
 .07   400 
 .071   394 
 .072   389 
 .073   384 
 .074   378 
 .075   373 
 .076   368 
 .077   364 
 .078   359 
 .079   354 
 .08   350 
 .081   346 
 .082   341 
 .083   337 
 .084   333 
 .085   329 
 .086   326 
 .087   322 
 .088   318 
 .089   315 
 .09   311 
 .091   308 
 .092   304 
 .093   301 
 .094   298 
 .095   295 
 .096   292 
 .097   289 
 .098   286 
 .099   283 
 .1   280 
 .13   215 
 .16   175 
 .19   147 
 .22   127 
 .25   112 
 .28   100 
 .31   90 
 .34   82 
 .37   76 
 .4   70 
 .43   65 
 .46   61 
 .49   57 
 .52   54 
 .55   51 
 .58   48 
 .61   46 
 .64   44 
 .67   42 
 .7   40 
 .73   38 
 .76   37 
 .79   35 
 .82   34 
 .85   33 
 .88   32 
 .91   31 
 .94   30 
 .97   29 
 1   28 
 1.03   27 
 1.06   26 
 1.09   26 
 1.12   25 
 1.15   24 
 1.18   24 
 1.21   23 
 1.24   23 
 1.27   22 
 1.3   22 
 1.33   21 
 1.36   21 
 1.39   20 
 1.42   20 
 1.45   19 
 1.48   19 
 1.51   19 
 1.54   18 
 1.57   18 
 1.6   18 
 1.63   17 
 1.66   17 
 1.69   17 
/
\endpicture$$

\noindent
\centerline{Figure 3. Plot of the confining potentials $W_1$ and $W_2$.
The dotted line is $\alpha_sC_F/r$.}

\end{document}